**Title**

Somatosensory prediction in premature neonates: iatrogenic pain experience increases repetition suppression and deviance detection of innocuous stimuli in a tactile oddball protocol

**Authors**

Anne-Lise Marais[a], Victoria Dumont[a], Marie Anquetil[b], Arnaud Mortier[c], Anne-Sophie Trentesaux[d], Nadège-Roche-Labarbe[a]

**Affiliations**

[a]Normandie Univ, UNICAEN, INSERM, COMETE, GIP CYCERON, Caen 14000, France

[b]Normandie Univ, UNICAEN, LPCN, Caen 14000, France

[c]Normandie Univ, UNICAEN, LMNO, Caen 14000, France

[d]CHU, Caen 14000, France

**Corresponding author**

nadege.roche@unicaen.fr

**Keys Words**

Sensory prediction, Neonatal cognition, Somatosensory cortex, Preterm neonates, Repetition suppression, Deviance detection, Mismatch response, Pain exposure, Neonatal pain, Premature birth, Electroencephalography, Event-related potentials

**Abbreviations**

cGA, corrected gestational age; EEG, electroencephalography; ERP, event-related potential; GA, gestational age; ISI, interstimulus interval; MMR, mismatch response; ND, neurodevelopmental disorders; NICU, neonatal intensive care unit; ROI, region of interest; RS, repetition suppression; SP, sensory prediction.

**Abstract**

Sensory prediction (SP) is a fundamental mechanism of perception that supports cognitive development. Atypical SP has been reported across multiple neurodevelopmental disorders (ND), suggesting it may constitute an early cross-syndromic marker. Premature birth is a major risk factor for ND, with risk increasing as gestational age (GA) at birth decreases. However, how perinatal risk factors shape the development of SP remains poorly understood. We do not know if untimely birth itself, or exposure to iatrogenic pain during neonatal intensive care, cause neurodevelopmental impairments. In this study, we first assessed whether SP can be detected in the brains of premature neonates at 35 weeks corrected GA using a tactile oddball-omission paradigm with EEG. We then examined the effects of the degree of prematurity and of the exposure to painful care procedures on neural indices of SP. Results demonstrate the presence of repetition suppression (RS) and a mismatch response (MMR) to deviant stimuli in the contralateral somatosensory cortex of premature neonates. The amplitude of these SP proxies was significantly affected by the number of painful procedures experienced since birth, independently of the effect of GA at birth. Contrary to our initial hypothesis that greater neurodevelopmental risk would be associated with less mature SP, infants with higher exposure to pain exhibited more robust indices of SP. These findings suggest that increased ex utero experience, even painful, is associated with accelerated maturation of predictive somatosensory processing. Longitudinal follow-up of participants at age 2 will explore how these early markers relate to developmental outcomes.

## 1    Introduction

Neurodevelopmental disorders (ND), such as autism spectrum disorders (ASD) or attention-deficit/hyperactivity disorder (ADHD), are characterized by impairments that affect multiple domains of development from early childhood and throughout the lifespan. These disorders affect approximately 1-17% of the population worldwide (Francés et al., 2022) and constitute a major societal and clinical challenge, because early identification is crucial for minimizing impairments and cost. Indeed, despite their variety of symptoms, ND share common risk factors, both genetic (Satterstrom et al., 2019) and environmental (Carlsson et al., 2021), suggesting that their origins lie in the prenatal or perinatal period. However, current screening methods rely on identifying behavioral anomalies months, if not years, after birth.

Premature birth, which affects more than one in ten births worldwide each year, is a major risk factor of ND and other persistent cognitive and behavioral impairments (Larsen et al., 2022). The lower is the gestational age (GA) at birth, higher is the risk of ND later in childhood (Pierrat et al., 2021). In addition, preterm-born children often show atypical sensory processing across multiple modalities, including both hypo- and hypersensitivities and impaired self-regulation, that resemble sensory profiles observed in ND (Bröring et al., 2017). Sensory profiles are more atypical when children were born at lower GA as well (Crozier et al., 2016; Machado et al., 2019). Sensory processing and self-regulation being considered the early foundation of developmental cascades, they appeared good avenues for research into early risk markers (Bradshaw et al., 2022).

A prominent hypothesis that emerged over the past decade posits that sensory and cognitive atypicality in children with ND may stem from early deficits in sensory prediction (SP) (Cruys et al., 2014; Lawson et al., 2014; Sinha et al., 2014). The brain's ability to exploit time intervals and rhythms from the sensory environment to build expectations about future events is a prerequisite of attention development, and deficits in time processing are indeed associated with attention deficits in infants (Colombo & Richman, 2002). In addition, SP in term neonates, assessed by the mismatch response to deviant sounds in an auditory oddball sequence, is proportional to behavioral inhibition at 1 year of age (Schwarzlose et al., 2023). For a review of predictive coding and SP in human cognitive research, with a developmental focus, see Marais & Roche-Labarbe (2025). Understanding how premature birth affects the development of SP mechanisms may thus provide critical insights into the developmental origins of ND and help identify markers for the early identification of neurodevelopmental vulnerability in infants.

Investigation of SP in preterm neonates is recent. Edalati et al. (2022) conducted an auditory oddball EEG study in preterm neonates at 30 to 34 weeks GA. The paradigm involved violations of rhythmic expectation, and results revealed a positive mismatch response (MMR) followed by a late positive component (LPC) at frontocentral electrodes, reminiscent of the P300 seen in older children when deviance orients attention. This suggests that even at this early age, the preterm brain engages in high-order prediction error processing, which is a sign that basic predictive mechanisms are in place before term. The auditory modality is most frequently used in SP protocols at all ages, but we prefer to focus on the somatosensory modality in preterm neonates. First, touch is the earliest sensory system to emerge during fetal development and serves as a foundation for the maturation of other modalities, playing a central role in cognitive and affective development (Bremner & Spence, 2017). Second, tactile processing and regulation of behavioral responses to touch are frequently impaired in ND and are a common feature of atypical sensory profiles (Cascio, 2010; Marco et al., 2011; Puts et al., 2014). Third, preterm neonates are exposed to repeated care procedures entailing aversive and/or painful tactile stimuli that durably affect their somatosensory processing (Maitre et al., 2017; Slater et al., 2010). These changes likely diffuse to cognitive processing of somatosentation such as prediction. Using a tactile omission paradigm where stimuli were presented at either fixed or jittered intervals, and optical imaging, we previously provided direct evidence of somatosensory SP in preterm neonates at 33 weeks corrected gestational age (cGA) (Dumont et al., 2022). Neonates exhibited suppression (RS) of brain activity in the somatosensory cortex (visible during unexpected omissions of the trial) in the highly predictable condition (fixed intervals), whereas in the less predictable condition (jittered intervals), a prediction-induced increase in brain activity appeared during omissions. These results showed top-down regulatory effects of SP based on experience with the temporal structure of the sensory environment, weeks before term.

In the present study, we aimed to extend this finding by quantifying the effects of GA at birth and exposure to iatrogenic pain on measures of SP. The goal was to evaluate the sensitivity of potential neonatal markers of ND risk to two major perinatal risk factors. We chose electroencephalography (EEG) for a robust measure of trial-level brain responses, and designed a unimodal oddball-omission paradigm to explore proxies of SP in the tactile modality. We used the repetition of frequent standard trials to quantify RS between the beginning and the end of the sequence. In the predictive coding framework, when incoming stimuli match predictions, RS is assumed to reflect successful prediction based on a reliable internal model of the environment (Grill-Spector et al., 2006; Grotheer & Kovács, 2016). Rare deviants among the standards (*i.e.*, oddball) allowed us to quantify the mismatch response to deviance: in the predictive coding framework, when inputs do not match top-down predictions, a bottom-up prediction error is generated in low-order cortices and fed back to update the internal model of the expected sensory event. In older participants, this results in the mismatch negativity (MMN), a negative deflection in EEG signals observed 200–300ms after stimulus onset (Näätänen, 2009), but in preterm neonates it can be positive and therefore, called the mismatch response (MMR) (Edalati et al., 2022). In addition to quantifying RS and the MMR to deviants, which are the most common proxies of SP described in the literature, we wanted to explore the response to omissions. Omission of an expected trial is thought to provide an opportunity to measure top-down prediction signals, without the confounding bottom-up processing activity (Chennu et al., 2016). We explored this prediction-induced activity during omissions. Finally, we explored the MMR to postomission trials. In pioneering works on infant SP using EEG, Nelson et al. (1990) found that the most visible sign of SP was not activity during the omitted trial, but a MMR to the standard trial immediately following an omission.

This study aimed to describe four proxies of sensory prediction (SP) in the tactile modality, in the brain of premature neonates at 35 weeks of cGA, in four regions of interest (ROI): somatosensory (low-order in the predictive coding framework) and frontal (high-order in theory, but not necessarily so at such a young age), in the contralateral and ipsilateral (to the stimulus) hemispheres. We hypothesized that changes would be significant in the contralateral somatosensory ROI, but not in the ipsilateral ROI. We did not have *a priori* hypotheses for the frontal areas. We then used the significant variables (RS and Deviant MMR) to analyze the effect of GA at birth and of the number of painful care procedures undergone since birth on their values. We hypothesized that a lower GA at birth and/or a higher number of painful procedures would impair sensory processing and cognitive development and thus lead to lower-amplitude SP proxies. Determining the link between perinatal risk factors and the magnitude of SP proxies is crucial for understanding the emergence of atypical trajectories and identifying markers for neonatal screening of ND.

## 2    Material and methods

### 2.1    Participants

We included 90 preterm neonates (43 females) born before 34 weeks + 6 days of GA in the Neonatal Intensive Care Unit (NICU) of the University Hospital of Caen, France. All participants were included after we obtained approval from their neonatologist regarding the infant's health and the parents' ability to provide informed consent. The study was approved by the ethics committee CPP Ile de France II, France, and pre-registered before inclusions with the Agence Nationale de Sécurité du Médicament et des produits de santé (ANSM, France) (ID RCB: 2020-A02117-32) and at the US National Institute of Health (NIH) registry of clinical trials (NCT04703010). Promotion and quality control were carried out by the Clinical Research Unit of the University Hospital of Caen, France.

The time of measurement was planned for the 35th week of cGA, with the approval of the participant's nurse. Exclusion criteria for the time of measurement were invasive respiratory assistance, neurological disease (intraventricular hemorrhage grade 3 or 4, periventricular leukomalacia, or major brain structure alteration assessed by transfontanellar ultrasound), suspected viral infection, patent bacterial infection (C-reactive protein concentration >20 mg/L), ongoing sedation (Fentanyl, Midazolam, Ketamine) for 48 hours preceding the day of the measurement.

Two infants were excluded before the measurement due to health deterioration, four because they were transferred to another hospital, and one due to high EEG impedances during acquisition that could not be improved, bringing the sample size to 83 (39 females). Gestational age at birth and the number of painful care procedures (all skin-invasive procedures, as well as intubation) since birth were recorded from the medical file *(Figure 1)*. Two participants did not have values for Pain because they had been transferred from another hospital, and we were not able the retrieve the data from the transferred files.

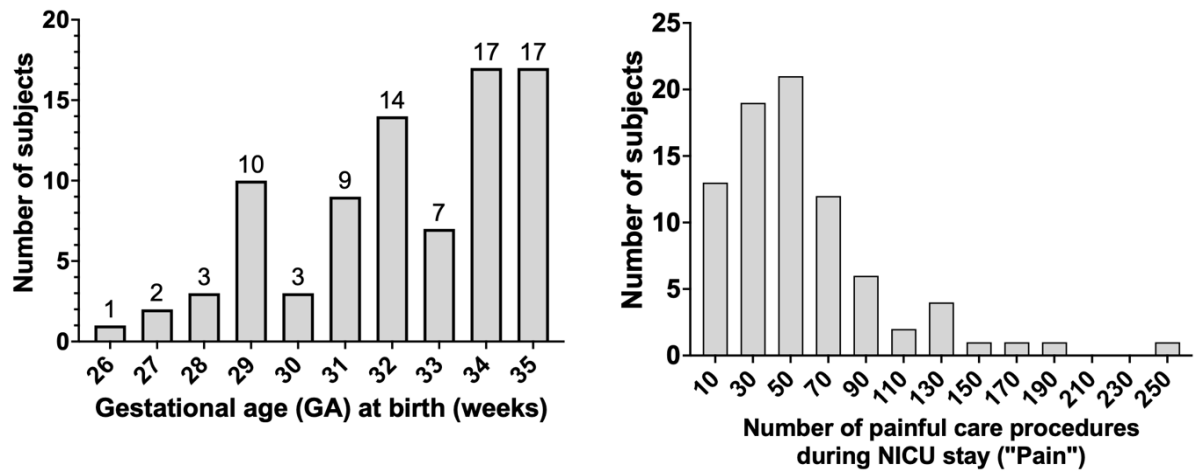

*Figure 1.* A. Distribution of the gestational age at birth in the final sample. B. Distribution of the number of painful procedures reported in the patient's file at the time of measurement.

### 2.2 Tactile stimulation protocol

Tactile stimuli were delivered by a custom-made vibratory matrix made of four vibrators in a row, enfolded in soft medical silicon (Caylar® SAS, Villebon-sur-Yvette, France), placed on the participant's forearm, and maintained using a tubular elastic net bandage *(Figure 2)*.

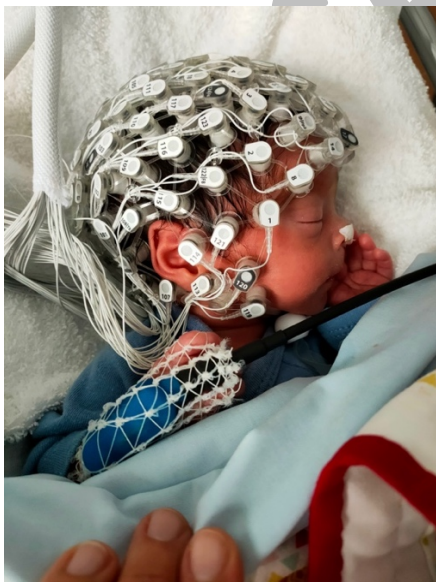

*Figure 2.* Study participant wearing the EEG net, and the vibrotactile stimulation matrix on the right forearm, during acquisition.

Vibrators were activated in turn, with an overlap, producing a 200ms tickling caress. The right forearm was stimulated when possible (absence of a catheter or bruise from previous medical procedures), but the left arm had to be used in 18 participants.

The stimulation sequence was designed with E-Prime® (Psychology Software Tools, Inc., Pittsburgh PA, USA). It provided 290 trials across three contiguous phases, lasting a total of 18 minutes. The first (Familiarization) and the last (Control) phases contained 40 standard trials. In between was the oddball-omission phase, composed of 30 contiguous blocks of seven pseudo-randomly ordered stimuli: five standards, one deviant (the direction of the stimulus was reversed), and one omission. Non-standard trials were never consecutive, and the directions of the standard and deviant trials were randomly counterbalanced between subjects. The interstimulus interval (ISI) was jittered between 3300 and 3700ms (average 3500, flat distribution). Omissions were always 3500ms apart from standards, thus generating a total 7200-ms interval. The standard trial following an omission will be called a Postomission trial thereafter because, although standard in its physical properties, it occurs after an omission that can be perceived as increased ISI and therefore may not be perceptually standard. In each block, one standard was randomly labeled as the "standard matched to deviant" (SMTD) trial, for subsequent statistical comparisons *(Figure 3)*.

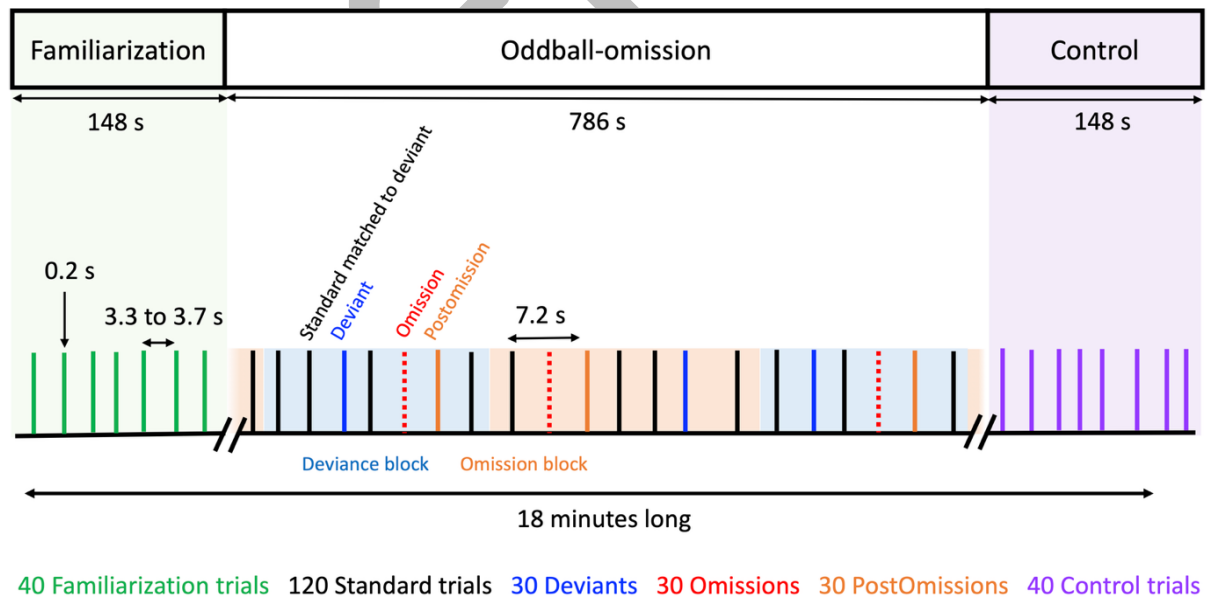

*Figure 3.* Schema of the stimulation sequence. Colors correspond to the ERPs by condition presented in the Results section.

### 2.3  Electroencephalography

*Acquisition*

Electroencephalography (EEG) was performed using a 128-channel system with saline-based electrode nets (Magstim® EGI, Eugene, OR, USA) *(Figure 2)* at a 1000 Hz sampling rate. Impedances were under 5 kΩ at the beginning of the measurement.

Measurements were performed during the 35th week of cGA in the infant's cradle or incubator, during the first sleep period following a feeding. In premature neonates, the first sleep cycle allows sensory and cognitive processing (Fifer et al., 2010). Parents were asked not to touch their children during the procedure. The protocol was stopped if the baby cried for a long period or needed medical care.

*Processing*

Raw files from the EGI system were processed in Matlab® (The Mathworks, Inc., Natick, MA, USA) using custom scripts (https://github.com/Roche-Labarbe/NEOPRENE_Project) and the EEGlab toolbox (Delorme & Makeig, 2004). We removed jaw and neck channels and down-sampled the data to 250 Hz before applying a bandpass filter (0.5–20 Hz), as well as a 50 Hz notch filter to eliminate line noise. We applied Artifact Subspace Reconstruction (ASR) to detect and correct bad channels and transient artifacts, then performed Independent Component Analysis (ICA). The resulting components are automatically labeled to identify non-neural artifacts, such as muscle or eye activity, and removed. After ICA, we re-referenced the EEG signals to the common average.

We segmented the data into epochs from 500ms before to 1s after trial onsets for Familiarization, Control, Deviant, SMTD, Omission, and Postomission trials.

To account for pre-onset neural activity associated with predictive processing, that could precede trial onset when the ISI were longer due to jittering, we applied a dynamic baseline correction window: this window was not locked to the trial onset $t_0$, but to the theoretical onset of prediction-induced activity, considering the shortest ISI in the sequence. For example, a trial preceded by a 3400ms ISI was corrected using the [-200 -100] values, a trial preceded by a 3500ms ISI was corrected using the [-300 -200] values, etc.

We then averaged epochs by condition to obtain event-related potentials (ERPs) and averaged the data from electrodes in four regions of interest (ROI): the left and right primary somatosensory cortices and the left and right frontal cortices *(Figure 4)*. Based on whether the subject was stimulated on the left or right forearm, we labeled each ROI as ipsilateral or contralateral to stimulation. For each subject, we removed regional ERPs that exceeded a variance of 60 or an amplitude of 40µV. We then calculated the proportion of trials retained in each ROI and kept only the ERP data with a proportion of at least 60% retained trials to calculate average ERPs for each ROI.

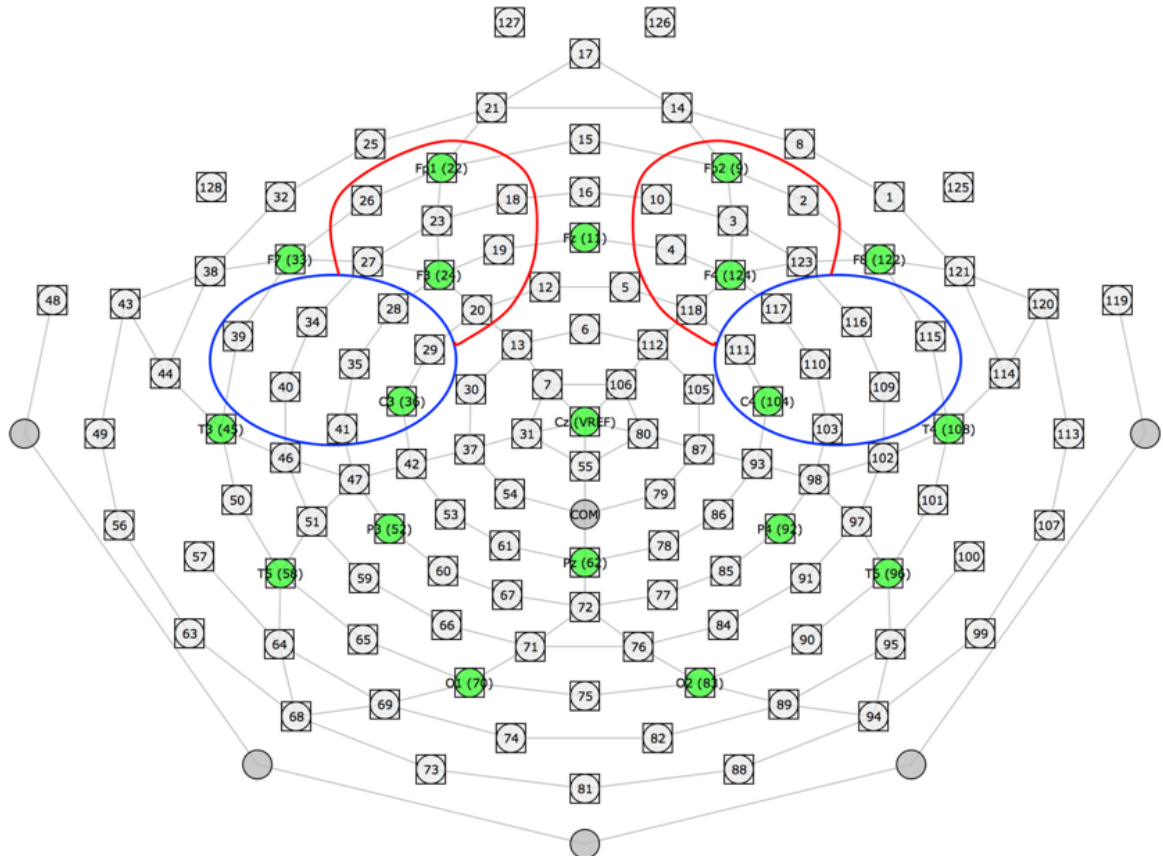

**Figure 4.** *Electrode groups constituting the regions of interest. Somatosensory ROIs are in blue, frontal ROIs are in red. View is from above the head, nose up. Green electrodes correspond to the 10-20 system of electrode placement.*

*Analysis*

RS was calculated by subtracting the N3 amplitude of Control trials from the N3 amplitude of Familiarization trials. The mismatch response to deviance, hereafter named DeviantMMR, was calculated by subtracting the N3 amplitude of SMTD trials from the N3 amplitude of Deviant trials. The mismatch response to Postomission trials (deviance resulting from an increased ISI between standard trials when a vibration was omitted, hereafter named PostomMMR) was calculated by subtracting the N3 amplitude of SMTD trials from the N3 amplitude of Postomission trials. For each of these conditions and each ROI, we calculated the amplitude of the N3 peak by averaging 50ms of data around the most negative value found between 450 and 650ms after trial onset of the "test" condition (window defined from grand averages of the Familiarization trials) and calculated the 50-ms average at the same latency on the corresponding control ERP.

We also analyzed the changes in activity occurring during Omissions, *i.e.*, prediction-induced activity. In the absence of *a priori* knowledge on such changes, we averaged the values between 3500 and 3700ms after the trial preceding the Omission, where the brain would theoretically expect a trial (marked 0 and 200ms on Omission figures, 0 being the theoretical onset of the expected trial). This value will be called Omission activity.

### 2.4   Statistics

We first compared the ERPs during tested conditions with their respective control trials. For the four ROIs, we calculated paired-samples two-tailed Student's t-tests between N3 amplitudes of Familiar *vs.* Control trials, Deviant *vs.* their SMTD trials, and Postomission *vs.* their SMTD trials, and we calculated one-sample two-tailed Student's *t*-tests between Omission activity and the theoretical value of 0. We adjusted *p*-values for multiple (16) tests using the Bonferroni correction.

The results of the previous analysis confirmed that an effect was only visible in contralateral ROIs and on the RS (Familiarization – Control) and DeviantMMR (Deviant – SMTD) markers (see section 3.1 Condition comparisons). Therefore, we fitted a random-intercept-only linear mixed-effects model with the values ("Marker value") of these two markers as the outcome (within-subject binary factor "Marker Type"), to determine their potential predictors. A second within-subject binary factor was the "Region" (Frontal – Somatosensory). Three covariates were included: GA, log(Pain), and the markers' "Standard value" (Familiarization, and SMTD, respectively) to control for an expected regression-to-the-mean (RTM) effect. Two interaction effects were added: Region × Marker Type, and GA × log(Pain). Subject was the only grouping factor.

<u>Primary model:</u> *Marker value ~ 1 + Region + Marker Type + Standard value + GA + Log(Pain) + Marker Type × Region + GA × Log(Pain) + (1|Subject).*

We opted for a "compound symmetry" residuals structure, "unstructured" implying a high risk of overfitting due to estimating many parameters with a relatively small sample size. Reported parameters were estimated by REML to limit bias, but we used ML estimations when performing model comparisons (see Robustness checks in Results). REML and ML are expected to give similar estimates given our sample size (Snijders & J., 2012) and we confirm that it is the case. The between-subject binary factors "Side" (stimulated on the right or left arm) and "Sex" were first added to the model only to check that they are irrelevant as planned, then removed (see Robustness checks in Results).

Statistics were performed using the open R interface jamovi ([https://www.jamovi.org/](https://www.jamovi.org/)), version 2.7.4.0. The GAMLj3 toolbox was used for linear mixed modelling in jamovi ([https://gamlj.github.io/](https://gamlj.github.io/)).

# 3    Results

## 3.1    Condition comparisons

Grand average ERPs are presented on *Figure 5*. All comparison statistics are reported in *Table 1*.

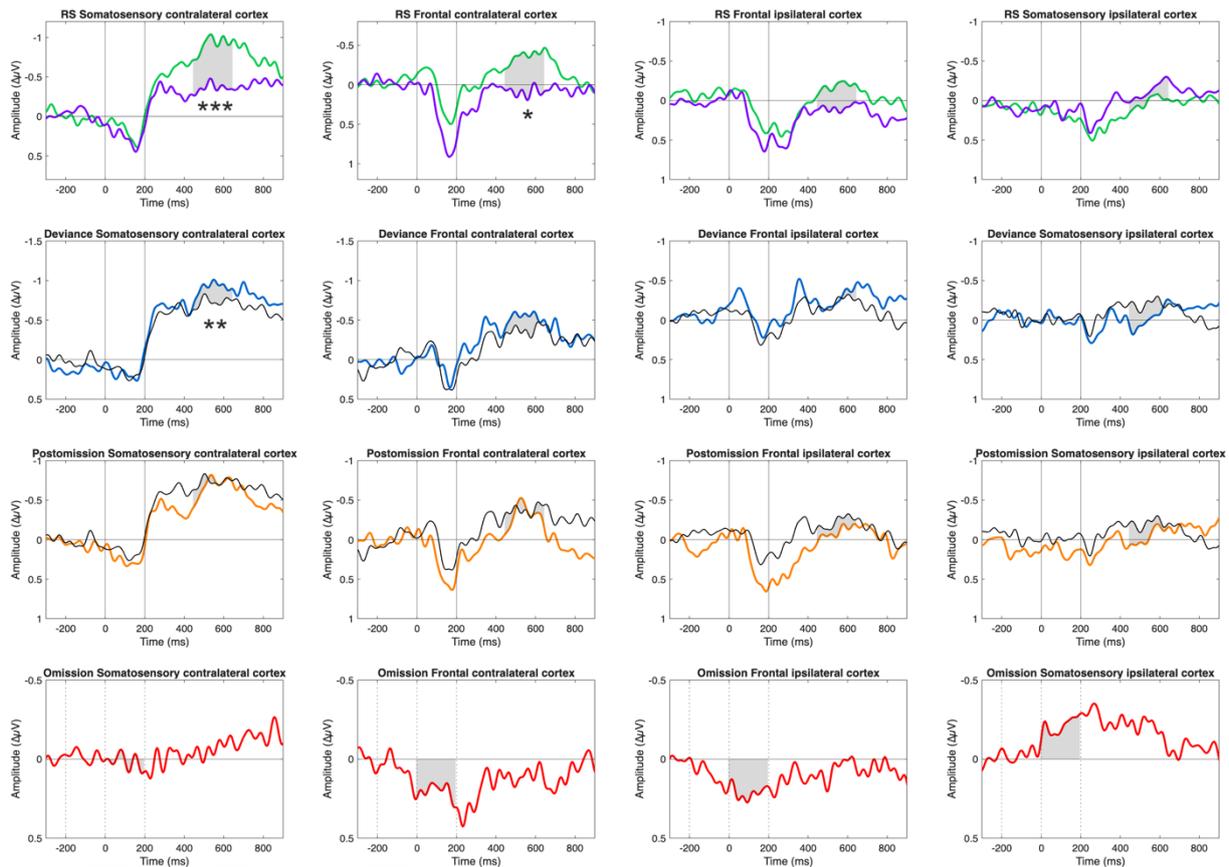

**Figure 5.** *Grand averages of ERPs by condition (in line) and ROI (in colums). Colors for each condition correspond to the colors on the stimulation sequence schema: purple for Familiarization phase Standards, green for Control phase Standards, blue for Deviants, orange for Postomissions, black for their respective matched standard trials, and red for Omissions. ROIs are represented from left to right: contralateral sensory, contralateral frontal, ipsilateral frontal, ipsilateral sensory. Significance is represented after Bonferroni correction for multiple comparisons.*

*Repetition suppression*

RS was highly significant in the contralateral somatosensory ROI before correction (*p*<0.0001, *t*=4.9, df=77) and remained highly significant after correction (*p*<0.0001). It was very significant in the contralateral frontal ROI before correction (*p*=0.002, *t*=3.2, df=62) and remained significant after correction (*p*=0.027). RS was significant in the ipsilateral frontal ROI before correction (*p*=0.032, *t*=2.2, df=66) but not after correction (*p*=0.514), and it was not significant in the ipsilateral somatosensory ROI (*p*=0.247, *t*=1.2, df=74). Note that degrees of freedom vary because not all 83 subjects had ERPs for all conditions, depending on bad segment rejection.

*Deviance response*

DeviantMMR was highly significant in the contralateral somatosensory ROI before correction (*p*=0.0002, *t*=3.8, df=78) and remained very significant after correction (*p*=0.004). It was significant in the ipsilateral somatosensory ROI before correction (*p*=0.041, *t*=2.1, df=75) but not after correction (*p*=0.649). It was not significant in the frontal ROIs.

*Postomission response*

We did not observe any significant difference between the Postomission trials and their respective SMTD trials.

*Changes during omissions*

Although low-amplitude Omission activity is visible on grand averages for three ROIs, it did not significantly differ from zero.

**Table 1.** *Statistical comparisons of all conditions and ROI.*

| Test # | Measure 1 | Measure 2 | Test | statistic | df | p-value | Bonferroni corrected (x16) |
|---|---|---|---|---|---|---|---|
| 1 | ContraFrontal_Fam | ContraFrontal_Control | | -3.185 | 62 | **0.0022627** | **0.0271522** |
| 2 | ContraS1_Fam | ContraS1_Control | | -4.900 | 77 | **0.0000052** | **0.0000829** |
| 3 | IpsiFrontal_Fam | IpsiFrontal_Control | | -2.189 | 66 | **0.0321361** | 0.5141780 |
| 4 | IpsiS1_Fam | IpsiS1_Control | | -1.167 | 74 | 0.2470814 | |
| 5 | ContraFrontal_Dev | ContraFrontal_SMTD | | -1.854 | 67 | 0.0680931 | |
| 6 | ContraS1_Dev | ContraS1_SMTD | paired-samples | -3.838 | 78 | **0.0002510** | **0.0040157** |
| 7 | IpsiFrontal_Dev | IpsiFrontal_SMTD | Student's t-test | -1.595 | 64 | 0.1157509 | |
| 8 | IpsiS1_Dev | IpsiS1_SMTD | $(H_a\ \mu1-\mu2 \neq 0)$ | -2.084 | 75 | **0.0405861** | 0.6493783 |
| 9 | ContraFrontal_Postom | ContraFrontal_SMTDPom | | -1.388 | 65 | 0.1698360 | |
| 10 | ContraS1_Postom | ContraS1_SMTDPom | | -0.666 | 77 | 0.5074738 | |
| 11 | IpsiFrontal_Postom | IpsiFrontal_SMTDPom | | -0.524 | 65 | 0.6020843 | |
| 12 | IpsiS1_Postom | IpsiS1_SMTDPom | | -1.186 | 75 | 0.2393503 | |
| 13 | ContraFrontal_Om | — | | 1456 | 69 | 0.2125754 | |
| 14 | ContraS1_Om | — | Wilcoxon's W | 1729 | 81 | 0.9006667 | |
| 15 | IpsiFrontal_Om | — | $(H_a\ \mu \neq 0)$ | 1482 | 71 | 0.3472388 | |
| 16 | IpsiS1_Om | — | | 1438 | 78 | 0.4892386 | |

### 3.2 Effect of the predictors

The Marker values were significantly larger in the Somatosensory region (estimated marginal mean M = -0.873, SE = 0.144) than in the Frontal region (M = -0.472, SE = 0.152, Mdiff = 0.401, SE = 0.158, CI95% = [0.089, 0.714], t(197) = 2.534, p=0.012).

There was no significant main effect of the Marker Type (F(1, 197) = 0.417, p = 0.519), and there was no significant Region × Marker Type interaction effect (F(1, 197) = 0.318, p = 0.573), indicating that RS and DeviantMMR behave similarly.

Regarding the covariates, we found a significant relationship with log(Pain) such that, after controlling for GA, a rise by 1 order of magnitude in the number of painful care procedures is associated with a decrease of the average Marker values by 0.942 units (b = 0.942, SE = 0.443, CI95% = [0.060, 1.825], t(76) = 2.126, p = 0.037). *Figure 6* presents regression lines of the marker values against log(Pain) for each brain region (in the contralateral hemisphere).

There was no significant effect of GA (b = 0.110, SE = 0.069, CI95% = [-0.028, 0.248], t(76) = 1.588, p = 0.116) and the log(Pain) × GA interaction effect was not significant either (F(1, 76) < 0.001, p = 0.996).

This suggests that despite the strong correlation (Bravais-Pearson's r=-0.79) between GA and the number of painful care procedures, only the latter is truly related to the markers' amplitudes. However, to clear any doubts related to a multicollinearity issue, we also fitted the model with either of the covariates log(Pain) or GA removed:

<u>Auxiliary model 1:</u> *Marker value ~ 1 + Region + Marker Type + Standard value + GA + Marker Type × Region + GA × Log(Pain) + (1|Subject).*

<u>Auxiliary model 2:</u> *Marker value ~ 1 + Region + Marker Type + Standard value + Log(Pain) + Marker Type × Region + GA × Log(Pain) + (1|Subject).*

In Auxiliary model 1 the main effect of GA is close to 0 and not significant (F(1, 78) = 0.010, p = 0.921). In Auxiliary model 2 the main effect of Log(Pain) is also not significant (b = 0.393, SE = 0.279, F(1, 78) = 1.981, p = 0.163).

This supports the idea that, despite non-significant, there is a slight effect of GA in the primary model.

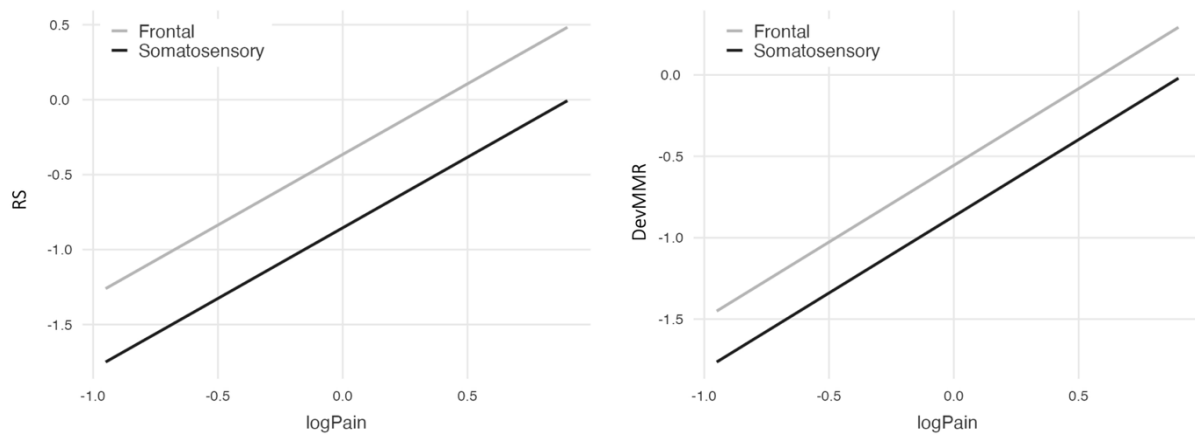

**Figure 6.** *Regression lines of the Marker values (RS on the left, DeviantMMR on the right) against log(Pain) for the contralateral frontal (gray) and contralateral somatosensory ROIs.*

Finally, there was a significant effect of the Standard values, which indicates a regression-to-the-mean (RTM) effect (b = -0.838, SE = 0.062, CI95% = [-0.716, -0.959], t(197) = 13.597, p<0.001).

The marginal $R^2$ of the model was 0.422, indicating that 42.2% of the variations in Marker values are explained by the fixed effects discussed above. We suspected that this value was largely driven by the RTM effect. Including the baseline (here the Standard value) as a covariate in a model with score changes as the outcome variable can yield biases in case of a strong RTM effect. Therefore, we also fitted the same model to the "Test values" (Control and Deviant) as a robustness check. The fixed-effect pattern was unchanged compared to the primary model, except for the unstandardized estimate of the Standard values covariate that was lower by approximately 1 unit (which is expected consequently to the algebraic shift between the two models). This confirms that the fixed effects found in the primary model are not artifacts driven by the RTM effect. The marginal $R^2$ of the <u>auxiliary model 3</u>: *Test value ~ 1 + Region + Marker Type + Standard value + GA + log(Pain) + Marker Type × Region + GA × log(Pain) + (1|Subject)*, however, is equal to 0.075, indicating that the effect size of the fixed effects discussed above is small to medium.

Assumption checks:

- Visual inspection of the Q-Q plot and residuals plot showed that the residuals are normally distributed and that the assumption of homoscedasticity is met.
- Inspection of the residuals-by-cluster plots showed normal boxplots with six exceptions that could be explained by missing data in several conditions.

Additional robustness checks:

- Adding Side (stimulation on the right or left arm) (ML ΔAIC = 1.96, ΔBIC = 5.61 in favor of the primary model, ΔR² < .001) or Sex (ML ΔAIC = 1.16, ΔBIC = 4.80 in favor of the primary model, ΔR² < 0.001) as between-subject binary factors has close to no impact on the analysis results.
- Using Unstructured residuals gives the same pattern of fixed effects, only with increased effect sizes.
- We fitted the same model to the measures from the ipsilateral hemisphere, in order to compare the primary model results to what is supposed to be unrelated activity. As expected, no fixed effects were significant in the ipsilateral model.

**4    Discussion**

This study measured proxies of somatosensory prediction at 35 weeks cGA in premature neonates of a wide range of GA at birth, and tested how GA at birth and the number of painful procedures undergone by the participant during hospital stay predicted these values. Using a tactile oddball-omission paradigm, we report that repetition suppression (RS) and a mismatch response to deviant stimuli (DeviantMMR) are both present in the contralateral somatosensory cortex of premature neonates (and in the contralateral frontal area to a lesser extent). These results provide evidence of sensory prediction in the tactile modality in premature neonates. In addition, we show a marked effect of the number of painful procedures undergone by the baby during hospital stay. This effect is distinct from the effect of GA at birth. Contrary to our initial hypothesis that greater neurodevelopmental risk would be associated with less mature predictive mechanisms, we found that the preterm infants in our cohort who had undergone more procedures exhibited the clearest signs of early predictive somatosensory processing. Lower GA at birth also increased the amplitudes of prediction markers, albeit to a lesser extent than pain exposure. Overall, we conclude that the increased *ex utero* exposure to somatosensory stimuli, particularly painful ones, is linked to more mature predictive markers; that is, both repetition suppression and the deviance response are larger.

The amplitude of the N3 component of ERPs was strongly reduced after repetitions of the Standard stimulus, whereas Deviant stimuli elicited an increased amplitude. These RS and mismatch effects indicate that a stable representation of the standard stimulus was formed, and used to expect the most likely input, according to the predictive coding framework – for a recent review of these effects, see Marais & Roche-Labarbe (2025). We thus confirm previous findings which showed that the preterm brain can form internal models and respond to violations of sensory expectations in the somatosensory modality (Dumont et al., 2022) and in the auditory modality (Edalati et al., 2022). In addition, we show that temporal uncertainty does not preclude RS. In our previous paper, we had observed RS only in the condition where ISI was fixed (*i.e.*, without any jitter) (Dumont et al., 2022). This discrepancy is likely due to the higher number of trials presented here, combined with a smaller jitter range resulting from smaller ISI and trial durations, all due to using EEG measures of neural activity as opposed to the slower neurovascular-based technique of DCS used in our previous work. Our results show the most significant RS in the contralateral sensory ROI, as expected, but RS is also significant in the frontal ROI, in line with accounts of RS as an active regulation mechanism involving higher-order cortical areas (Nordt et al., 2016; Summerfield et al., 2008). On the contrary, the Deviance MM response was only significant in the somatosensory ROI. In older populations, when the deviant trial is salient or when the subject's attention is directed towards the stimuli, the MM response in sensory areas is associated with frontal components, typically the P3, on ERPs (Riggins & Scott, 2020). Preterm neonates may be so young that the attention-orienting effect of deviance is not present, or so small that it is not measurable. Alternately, the absence of frontal activity after Deviants may reflect the stimulus' low relevance: we used a unimodal protocol with no complexity in the temporal structure and the tactile input was not social or otherwise salient. Further studies manipulating the relevance and salience of the stimulus, for example using social touch, or involving rewarding associations inspired from audio-visual paradigms that have been used in older infants (Emberson et al., 2015), will be necessary to determine if a frontal activity can be elicited in preterm neonates by deviant trials.

We did not observe any mismatch response to the Postomission trials. This is consistent with our previous study using an omission protocol: when ISI were jittered, we had not observed any difference between neurovascular responses to Postomission and Standard trials either, even though neurovascular responses to Omissions were present (Dumont et al., 2022). However, this was in contradiction with the results of a pioneering study using EEG and a visual omission protocol in 6-month-old infants (Nelson et al., 1990). The authors had found that when the stimulus was not presented following the auditory cue in an audio-visual association sequence, the brain responded with increased activity to stimuli presented after an omission, but not to the omission itself. There may be a discrepancy between SP based on temporal structure *vs.* on associative learning, with prediction signals coming from high-order areas *vs.* other sensory areas, respectively. Alternately, preterm neonates may not process omissions in the same manner as older infants. Further studies using bimodal stimulation protocols in neonates at various ages are necessary to untangle maturational effects from differences in underlying cognitive processing of PS depending on the paradigm.

Although brain activity during Omissions was not significantly different from 0, low-amplitude positive changes appear on grand averages of both frontal ROI, while a negative change is visible on the ipsilateral sensory ROI. Prediction-induced brain activity during omission of an expected stimulus has been reported in only a few studies, mostly in the auditory modality. In adults, Wacongne et al. (2011) described negative responses to omissions similar to the ones evoked by a presented stimulus. In newborns, the omission of the downbeat in a drums sequence was also associated with a negative wave, approximately 200ms after the expected onset (Winkler et al., 2009). In the visual domain, omissions were not associated with EEG changes in the experiment by Nelson et al. (1990), but Emberson et al. (2015) reported neurovascular changes using fNIRS and a similar audio-visual paradigm. This suggests that, rather than a lack of neuronal activity during omissions, the visual cortex of infants produced potentials that were too tangential to the scalp or spread in multiple directions that cancel out at the scalp level. In our neurovascular study, an increase in activity was visible when somatosensory stimuli were omitted in the jittered sequence (Dumont et al., 2022), suggesting that the low-amplitude changes we observed in the present study may also stem from neuronal activity that is too tangential or diffuse in the cortex to produce a significant ERP. The jittering, although useful to keep cognitive processes engaged through the sequence, constitutes an additional difficulty when analyzing Omission trials because it induces an uncertainty in the expected trial onset. As a result, prediction signals may arise before the hypothetical onset that is arbitrarily placed in the middle of the jittering range, up to 200ms before in our case. This may explain why the Omissions grand averages seem to start diverging from zero around -100ms. In order to better understand the prediction-induced activity during omissions, we may benefit from reanalyzing our data using a time-frequency approach. For future studies of pre-term SP, bimodal association paradigms would allow accurate timing of omissions while maintaining some interest in the input, but other relevant stimuli such as social touch could also encourage SP processes even with a simple temporal structure.

This study demonstrates for the first time that SP, and/or the regulation of brain activity in various conditions we measure as its proxies, depend on *ex utero* sensory exposure. When designing this study, we hypothesized that the more premature infants would be born and/or the more pain they had experienced, the less RS and MM they would exhibit. This hypothesis was based on the premise that such infants have a less favorable neurodevelopmental prognostic, suggesting less effective core cognitive mechanisms. However, our results provide evidence of the contrary in a sample where patent neurological insults were excluded, and suggest that the extent of extrauterine sensory experience is decisive in the absence of major brain damage even when experienced earlier than normal, and even when aversive. This accelerated maturation due to premature exposure is consistent with studies of simple ERPs (for a review and recent results in different sensory modalities, see Mellado et al. (2022). Using somatosensory ERPs, Maitre et al. (2017) reported that preterm neonates of various GA at birth, measured between 35 and 43 weeks cGA, exhibited reduced ERP amplitude compared to their term-born peers when they had been exposed to a higher proportion of painful stimuli than average. RS may be adaptive in the short-term, allowing preterm neonates to filter the dystimulating NICU environment, preserving cognitive and physiological resources during their hospitalization. However, it indicates the sharpening of mental models of the NICU environment through interactive specialization (Johnson, 2011). Even if lower GA at birth had a small effect, indicating that broader sensory inputs than just the noxious procedures were involved in the maturation of predictive mechanisms, pain experience had the largest effect in our data. Core cognitive processes built on such aberrant inputs are likely to compromise long-term neurodevelopment. Indeed, when the infant is discharged from the hospital, her perceptions will be biased towards non-social, irrelevant, or aversive inputs. Even without neurological damage, the child will be at risk of lasting atypical sensory processing of the typical home and family environment, hindering subsequent adaptive learning.

Further studies are necessary to explore specifically the role of pain on somatosensory prediction, using dedicated data collection methods. The number of painful procedures reported in the patient's charts, that we defined as breaking of the skin or invasive procedure, is an imperfect proxy for nociceptive experience. It depends on the accuracy and completeness of chart filling by clinicians. Unsuccessful attempts at procedures (*e.g.*, multiple venipuncture attempts) are not recorded, and for infants transferred into our unit within the first days of life, early data from the referring hospital are often incomplete or unavailable due to the lack of standardized electronic record systems.

Maitre et al. (2017) also reported that a higher exposure to positive tactile stimuli (social and not care-related) was associated with increased amplitudes of tactile ERPs. Positive tactile experiences may contribute to prevent the effects of painful experiences and thus promote sensory processing, by which learning will occur through the formation of predictive sensory models. The more positive experiences, especially when such stimulation is varied, multisensory and provided consistently, the more internal models will be ecologically relevant after discharge from the hospital and promote optimal neurodevelopment. Therefore, it will be important to quantify these events in future works, notably skin-to-skin contact for example through kangaroo mother care (KMC) which provides infants with contingent, behaviorally relevant tactile stimulation, and may alleviate deleterious aspects of the NICU experience (Kostandy & Ludington-Hoe, 2019; Minotti et al., 2025). Unfortunately, in the present study we could not quantify skin-to-skin exposure because this information is rarely documented in clinical records. A systematic data acquisition would be necessary to characterize each infant's tactile environment and to assess the role of the various aspects in modulating SP. However, it is interesting to mention that the University Hospital of Caen, where this study was conducted, strongly promotes KMC. Even though we cannot quantify it, many of our subjects benefited from extended and/or frequent KMC sessions, which could also contribute to SP measures increasing in amplitude with post-natal experience.

To fully describe the development of SP in premature neonates, longitudinal measurements would be ideal to follow the effect of post-natal experience on ERPs at different cGA. In short, future efforts should combine longitudinal measures with a systematic acquisition of all tactile interactions, using bimodal and/or behaviorally relevant stimuli. Finally, including infants with neurological injury would also inform on the relative weights of neurological impact of premature birth *vs.* post-natal *ex utero* exposure impact, at equivalent cGA. This many factors will require a larger sample size, but identifying early mechanisms and markers of atypical cognitive development in the NICU holds promise for early detection of infants at risk for later neurodevelopmental impairments. Given the fundamental role of early tactile experience in brain organization, criteria allowing to quantify the effect of tactile interventions (such as KMC or massage-therapy) on neurodevelopment is also an important perspective of this work.

We are currently conducting a longitudinal follow-up study of the participants of this study at 2 years of chronological age. A global evaluation of neurodevelopment is performed, including cognitive, motor, and socio-emotional domains and a reevaluation of sensory prediction. This ongoing work will allow us to investigate whether the presence or quality of neonatal SP is predictive of later developmental outcomes, and whether tactile prediction could serve as an early, non-invasive biomarker for identifying infants at risk for ND. Ultimately, such biomarkers could guide early interventions tailored to individual developmental profiles.

*Conclusion*

In this study, we show that predictive processing mechanisms can be measured with EEG before the term in premature neonates, and that infants who had experienced more painful care procedures relative to their gestational age at birth exhibited the largest amplitudes of the markers, namely repetition suppression and the mismatch response to deviance. These findings reveal neurophysiological traces of early adaptive plasticity in the context of developmental vulnerability and underscore the potential of predictive coding measures as early biomarkers of atypical neurodevelopmental trajectories. By examining how preterm neonates respond to the temporal structure of tactile stimulation, we may gain insight into the early maturation of statistical learning, expectation, and attention regulation. Follow-up measures at two years of age are ongoing to determine the predictive validity of these early EEG markers for later neurodevelopmental outcome.

## Funding

This research was supported by the NEOPRENE grant #ANR-19-CE37-0015 from the Agence Nationale de la Recherche (www.anr.fr) to N.R.L., a grant from the Fondation Perce-Neige (www.perce-neige.org) to N.R.L., and a Ph.D. grant from the Université de Caen Normandie to A.L.M.

## CRediT authorship contribution statement

A.L.M. designed and coded the protocol, acquired the data, analyzed the data, and wrote the manuscript. V.D. designed the protocol, acquired the data, analyzed the data, and wrote the manuscript. M.A. designed the protocol, and acquired the data. A.M. designed the linear mixed model and wrote the corresponding results. A.S.T. was responsible for patient clinical evaluation and inclusion. N.R.-L. designed the study, obtained funding, acquired the data, analyzed the data, and revised the manuscript.

## Declaration of Competing Interest

The authors declare that they have no known competing financial interests or personal relationships that could have appeared to influence the work reported in this paper.

## Data Availability

Anonymized data will be available upon request to Dr. Nadege Roche-Labarbe by email at nadege.roche@unicaen.fr and after a collaboration agreement has been approved by the University of Caen Normandy, France, and by the Caen University Hospital, France, co-owners of the data, with the requesting institution.

**Acknowledgments**

We are immensely grateful to all the babies and their parents for their participation, and the nurses, physicians, and staff in the Neonatology Unit at the Caen University Hospital for their unwavering help and support.